\documentclass{amsart}

\usepackage[utf8]{inputenc}
\usepackage{hyperref}
\usepackage{natbib}
\usepackage{graphicx}
\usepackage{subfiles}
\usepackage{float}

\newcommand{\notarxiv}[1]{}
\newcommand{\arxiv}[1]{#1}

\title{torchtree: flexible phylogenetic model development and inference using PyTorch}

\usepackage{amsmath, amsfonts, bm}
\DeclareMathOperator*{\argmin}{argmin}

\newcommand{\BF}{\operatorname{BF}}
\newcommand{\KL}{\operatorname{KL}}

\newcommand{\torchtree}{\texttt{torchtree}}
\newcommand{\phylostan}{\texttt{phylostan}}
\newcommand{\bito}{\texttt{bito}}
\newcommand{\physher}{\texttt{physher}}
\newcommand{\BEAGLE}{\texttt{BEAGLE}}
\newcommand{\BEAST}{\texttt{BEAST}}

\newcommand{\beginsupplement}{%
        \setcounter{table}{0}
        \renewcommand{\thetable}{S\arabic{table}}%
        \setcounter{figure}{0}
        \renewcommand{\thefigure}{S\arabic{figure}}%
}

\begin{document}

\author[Fourment et al.]{
Mathieu Fourment,$^{\ast,1}$
Matthew Macaulay,$^{1}$
Christiaan J Swanepoel,$^{2,3}$
Xiang Ji,$^{5}$
Marc A Suchard,$^{6,7,8}$
Frederick A Matsen IV$^{4,9,10,11}$
}

\maketitle

\medskip
\noindent{\bf Corresponding authors:}
Mathieu Fourment, Australian Institute for Microbiology and Infection, University of Technology Sydney, Ultimo NSW, Australia, E-mail: mathieu.fourment@uts.edu.au

\noindent{\small\it
 \address{
$^{1}$Australian Institute for Microbiology and Infection, University of Technology Sydney, Ultimo NSW, Australia;\\
$^{2}$Centre for Computational Evolution, The University of Auckland, Auckland, New Zealand;\\
$^{3}$School of Computer Science, The University of Auckland, Auckland, New Zealand, 1010;\\
$^{4}$Public Health Sciences Division, Fred Hutchinson Cancer Research Center, Seattle, Washington, USA;\\
$^{5}$Department of Mathematics, Tulane University, New Orleans, USA;\\
$^{6}$Department of Human Genetics, University of California, Los Angeles, USA;\\
$^{7}$Department of Computational Medicine, University of California, Los Angeles, USA;\\
$^{8}$Department of Biostatistics, University of California, Los Angeles, USA;\\
$^{9}$Department of Statistics, University of Washington, Seattle, USA;\\
$^{10}$Department of Genome Sciences, University of Washington, Seattle, USA;\\
$^{11}$Howard Hughes Medical Institute, Fred Hutchinson Cancer Research Center, Seattle, Washington, USA;\\
}
}

\begin{abstract}
Bayesian inference has predominantly relied on the Markov chain Monte Carlo (MCMC) algorithm for many years.
However, MCMC is computationally laborious, especially for complex phylogenetic models of time trees.
This bottleneck has led to the search for alternatives, such as variational Bayes, which can scale better to large datasets.
In this paper, we introduce torchtree, a framework written in Python that allows developers to easily implement rich phylogenetic models and algorithms using a fixed tree topology.
One can either use automatic differentiation, or leverage torchtree's plug-in system to compute gradients analytically for model components for which automatic differentiation is slow.
We demonstrate that the torchtree variational inference framework performs similarly to BEAST in terms of speed and approximation accuracy.
Furthermore, we explore the use of the forward KL divergence as an optimizing criterion for variational inference, which can handle discontinuous and non-differentiable models.
Our experiments show that inference using the forward KL divergence tends to be faster per iteration compared to the evidence lower bound (ELBO) criterion, although the ELBO-based inference may converge faster in some cases.
Overall, torchtree provides a flexible and efficient framework for phylogenetic model development and inference using PyTorch.
\end{abstract}

\arxiv{
}

\section*{Introduction}
Markov chain Monte Carlo (MCMC) has been a staple for Bayesian inference in phylogenetics over the last twenty years.
It is widely used \citep{suchard2018bayesian,ronquist2012mrbayes} and is considered the gold standard because once converged, it samples the posterior distribution exactly (although with autocorrelation).
However, MCMC is best suited to smaller data sets because it is computationally laborious, especially for complex models centered around time trees.
Due to computational difficulties associated with MCMC, researchers with large data sets often opt for non-Bayesian methods based on maximum likelihood \citep{Sagulenko2018-bb}, parsimony \citep{turakhia2021usher} and other heuristics \citep{to2016lsd,tamura2012estimating}, although these methods are not amenable to complex model inference and, unlike with Bayesian inference, deriving and interpreting confidence intervals can be tricky.

Variational Bayes (VB), also called variational inference, is an alternative approach for Bayesian inference that can scale better to large datasets \citep{jordan1999introduction}.
VB uses optimization to find the closest approximation to the posterior from a family of densities.
Consequently, it tends to be faster than MCMC, although it is not guaranteed to provide an exact representation of the posterior.
One can choose from several criteria to define closeness but the Kullback-Leibler (KL) divergence is the most common.

The optimization process in VB algorithms, such as the automatic differentiation variational inference (ADVI) algorithm \citep{Kucukelbir2017-cr}, require the gradient of the posterior with respect to the model parameters.
If the model is complex, computing this gradient efficiently is especially important, either analytically or with automatic differentiation.
Automatic differentiation libraries such as PyTorch \citep{Paszke2019-di}, Tensorflow \citep{Abadi2016-xh}, stan-math \citep{Carpenter2015-math} and JAX \citep{Bradbury2018-jax} calculate these gradients for the optimizer.
Although these libraries are all Python-based, except stan-math, they all use C++ under the hood to speed up calculations.

\citet{Fourment2019-fh} introduced \phylostan, which was the first package that used automatic differentiation to approximate phylogenetic models using variational inference and Hamiltonian Monte Carlo (HMC) \citep{Neal2011-yo}.
This Python program generates phylogenetic models in the Stan language which are fed to the Stan program \citep{Carpenter2017-ui}.
It works with a single fixed topology with a wide range of phylogenetic models.
While creating new models is straightforward thanks to the ease of learning the Stan language, there are certain drawbacks to consider.
For example it lacks flexibility for extending the code base and it is slow compared to pure C/C++ code \citep{Fourment2019-fh} and, in some cases, to some Python libraries \citep{fourment2023automatic}.
In the context of variational inference, it can only optimize the evidence lower bound (ELBO) using either meanfield or fullrank variational distributions using the automatic differentiation variational inference framework \citep{Kucukelbir2017-cr}.
Providing starting values to the variational distribution is difficult and extending the code base is tedious as it requires a deep understanding of the C++ code base.
% looks like we can provide the mean of the variational distribution but not the variance

Variational Bayes has garnered increasing attention lately, as evidenced by a surge in related research papers~\citep{Dang2019-wq,Liu2021-lw,Moretti2021-ah,Ki2022-nt,Koptagel2022-bk,Zhang2022-tr,swanepoel2022treeflow}.
A major difficulty in applying VB to phylogenetic model is dealing with the topology, the discrete component of the model.

It is tempting to implement a phylogenetic package for VB entirely in a fast language such as C.
Early studies \citep{Fourment2020-qw,Fourment2019-fh,fourment2023automatic} showed significant improvements in terms of computational efficiency with such software at the cost of implementing the gradient of every model component.
Specifically, our previous benchmarking results~\citep{fourment2023automatic} suggested a strategy in which the computationally expensive parts of gradient computation (such as the phylogenetic likelihood) are computed with specialized algorithms~\citep{Ji2020-pp}, while less expensive model gradient calculations are performed using automatic differentiation.

We propose \torchtree, a framework written in Python, a language widely used by researchers, that allows developers to easily implement phylogenetic models and algorithms.
Inference is carried out under the assumption of a fixed tree topology, a common consideration in fast programs designed for large-scale phylodynamic analysis \citep{hadfield2018nexstrain,Sagulenko2018-bb,to2016lsd}.
Although this framework is heavily geared towards variational inference, other algorithms such as MCMC, HMC, and maximum \textit{a posteriori} inference are implemented.
Although \torchtree\ is already significantly faster than \phylostan\ when using automatic differentiation, we provide several plug-ins supplying analytic gradients that significantly improve its speed as shown in other studies \citep{Fourment2019-fh,fourment2023automatic}.

We find that ELBO-based variational inference (which uses a gradient) performs poorly on skygrid coalescent models due to the discontinuities present in the piecewise-constant population size function.
By using the forward KL divergence as the optimizing criterion, we can circumvent this continuity issue, as it does not require the model to be continuous or differentiable.
Our results also indicate that inference using the forward KL divergence tends to be faster per iteration compared to those using the ELBO, however in some cases the ELBO-based inference converges faster than the forward KL-based method.
In some cases, the approximation with the forward KL was just as accurate as the ELBO-based approximation.
We also propose a piecewise-linear adaptation of the skygrid which we call the \emph{skyglide}, and show that even though this model is not differentiable, it performs well under ELBO-based inference.

\section*{Variational inference}
Variational inference seeks to minimize a measure $\mu$ between the posterior and a simpler variational distribution chosen from a family of distributions.
For phylogenetic inference, the posterior probability of a tree's continuous parameters $\bm{z} \in Z$ conditioned on a sequence alignment $D$ with a fixed topology $\tau$ is $p(\bm{z}|D, \tau)$.
The variational distribution $q(\bm{z}; \bm{\phi})$ is defined over the tree's continuous variables and is parameterized by $\bm{\phi}$.
The objective is to find the optimal variational approximation from a family of functions $q(\bm{z}) \in \mathcal{Q}$:

$$
q^{*}(\bm{z}) = \argmin_{q(\bm{z}; \bm{\phi}) \in \mathcal{Q}} \mu (q(\bm{z}; \bm{\phi}) || p(\bm{z}|D, \tau)).
$$

The solution found through optimization $q^{*}$ then serves as an approximation for the posterior distribution.
Typically, practitioners use the backward Kullback-Leibler divergence for the measure between two distributions $\mu(q, p) = \KL(q||p)$~\citep{Blei2017-fs}.
It measures the amount of information lost by using the approximation $\KL(q||p) = \mathbb{E}[\log q(z; \bm{\phi})] - \mathbb{E} [\log p(z| D, \tau)]$.
Minimizing KL-divergence is equivalent to maximizing a lower bound of the evidence called the ELBO:

$$
\mathcal{L}(q) = \mathbb{E}[\log(p(\bm{z}, D | \tau))] - \mathbb{E}[\log(q(\bm{z}; \bm{\phi}))],
$$
where the expectations are taken with respect to the variational distribution $q$.
This comes by expanding the second term of the KL-divergence and dropping the intractable quantity $p(D, \tau)$, which is a constant and makes computing the ELBO relatively inexpensive.

Another choice of measure is the forward KL-divergence
$$\KL(p||q) = \mathbb{E}_{p(z| D, \tau)}[\log p(z| D, \tau) - \log q(z; \bm{\phi})].$$
This version reflects the ``intent'' of the KL-divergence, with the ``ground truth'' being the first term.
Despite its computational intractability, its mass-covering behavior does not result in underdispersed approximations of the distribution of interest \citep{naesseth2020markovian,jerfel2021variational}.
This is in contrast to the mode-seeking behavior associated with $\KL(q||p)$ which tends to underestimate the variance of the posterior (See Figure 1 in \citep{jerfel2021variational} for a visual explanation).

However, directly minimizing this measure poses challenges as it necessitates sampling from the posterior distribution.
We can use a self-normalized importance sampling (SNIS) gradient estimator~\citep{bornschein2015reweighted,jerfel2021variational} to estimate this measure and its gradient.
The importance sampling estimate of $\KL(p||q)$ using the instrument distribution $q$ is

\begin{align*}
\KL(p||q) & = \mathbb{E}_{p(z| D, \tau)}\log \left[\frac{p(z| D, \tau)}{q(z; \bm{\phi})}\right]\\
 & = \mathbb{E}_{q(z; \bm{\phi})} \left[\frac{p(z| D, \tau)}{q(z; \bm{\phi})}  \log \frac{p(z| D, \tau)}{q(z; \bm{\phi})} \right] \\
 & = \frac{\mathbb{E}_{q(z; \bm{\phi})} \left[\frac{p(z, D, \tau)}{q(z; \bm{\phi})}  \log \frac{p(z| D, \tau)}{q(z; \bm{\phi})} \right]}{\mathbb{E}_{q(z; \bm{\phi})} \left[\frac{p(z, D, \tau)}{q(z; \bm{\phi})} \right]} \\
 & \approx \sum_{s=1}^S \log\left(\frac{p(\tilde{z}_s| D, \tau)}{q(\tilde{z}_s ; \bm{\phi})}\right) w_s
\end{align*}

where $\tilde{z}_s \sim q(z; \bm{\phi})$ and

$$w_s = \frac{p(\tilde{z}_s, D, \tau)}{ q(\tilde{z}_s; \bm{\phi})} \Big/ \sum_{i=1}^N \frac{p(\tilde{z}_i, D, \tau)}{q(\tilde{z}_i; \bm{\phi})}.$$

An importance sampling estimate of the gradient of the $\KL(p||q)$ divergence can be computed similarly as follows.
Notice that $\mathbb{E}_{p(z| D, \tau)}[\log p(z| D, \tau)]$ does not depend on $\phi$, therefore minimizing $\KL(p||q)$ is equivalent to minimizing the cross entropy $L_{\KL}(\phi)$ with respect to variational parameters,

$$L_{\KL}(\phi) = \mathbb{E}_{p(z| D, \tau)} [\log q(z; \bm{\phi})].$$

Thus we can first take the gradient and then use the SNIS estimator, giving:

$$\nabla L_{\KL}(\phi) = -\sum_{s=1}^S w_s \nabla\log q(\tilde{z}_s ; \bm{\phi}) , \quad \tilde{z}_s \sim q(z; \bm{\phi}).$$

Another key advantage of using $\KL(p||q)$ is that the posterior does not have to be differentiable.
Opting out of gradient calculation can enhance computational efficiency, as the computational expense associated with gradient calculations is typically high.
In practice, employing the variational distribution as the importance distribution in the SNIS algorithm typically reduces the approximation variance.

\section*{Implementation}
\torchtree\ implements a variety of Bayesian inference techniques including maximum \textit{a posteriori} optimization, variational inference, and Hamiltonian Monte Carlo.
It also provides out of the box extensions to variational inference including multi-sampling, normalizing flows and various divergence measures.
It implements a wide range of phylogenetic models (Table~\ref{tab:models}), allowing investigating complex phylodynamic questions.
To demonstrate this functionality, we introduce some implementation details.

\torchtree\ is implemented in Python and uses PyTorch to leverage automatic differentiation.
Its design is inspired by the \BEAST\ packages \citep{suchard2018bayesian,bouckaert2019beast}, specifically the object structure, the plugin architecture, and the file formats.
In \torchtree\ the specification of model and algorithm parameters are specified through JSON files while \BEAST\ uses the XML format. 
Much like BEAST2, \torchtree\ provides a simple framework for creating plug-ins without modifying the existing code base.

In some cases some functions can be more efficiently written directly in C++.
Indeed, \citet{Fourment2019-fh,fourment2023automatic} showed that analytical derivatives generally outperform those derived through automatic differentiation.
Mixing automatic differentiation and C++ code is possible with custom C++ extensions provided by PyTorch.
We describe later in this section \torchtree\ plug-ins that make use of these custom extensions.

A simple command-line interface is provided in order to build configuration files.
However, the user will need to edit it manually in order to adjust parameters such as hyperprior parameters.
%(unlike BEASTs which have BEAUTI but lot of work to do that, I played a bit with electron to have an html based interface similar to BEAUTI but it is bloody hard to implement anything).
%Documentation is provided using Python docstrings in order to understand the building blocks.

\torchtree\ is open source software, available on GitHub at \url{https://github.com/4ment/torchtree}, along with compiled API documentation.

\subsection*{Phylogenetic models}

\begin{table}
\centering
\begin{tabular}{ |c|c| }
 \hline
 Type & Model \\
 \hline
Nucleotide substitution model & Standard reversible models and SRD06 \\
Amino acid substitution model & WAG and LG \\
Codon model& GY94 \\
Rate heterogeneity across site & Proportion of invariant sites, Weibull \\
Phylogeography & Discrete \\
Birth death & Constant, BDSKY \\
Coalescent & Constant, exponential, skyride, skygrid, skyglide \\
Clock prior & Clock-free, strict, autocorrelated, uncorrelated\\
 \hline
\end{tabular}
\caption{Base models implemented in \torchtree.}
\label{tab:models}
\end{table}

\subsection*{Plug-ins}

\subsubsection*{torchtree-physher}
The \texttt{torchtree-physher} plug-in uses \physher\ \citep{Fourment2014-sa}, a C-based program, to efficiently evaluate several likelihood functions and their gradients (Table~\ref{tab:physher-models}).
Every model implemented in \texttt{torchtree-physher} is analytically differentiable, except the gamma site model for which the gradient is approximated numerically.
The Weibull site model, suggested by \citet{Fourment2019-fh}, serves as an alternative to the widely used gamma site model \citep{yang1994maximum}.
It is favored for its closed-form inverse cumulative distribution function (CDF), allowing straightforward analytical gradient calculation.
The derivatives with respect to the branch lengths are efficiently calculated using a linear-time algorithm \citep{Fourment2020-qw}, resulting in a substantial speed boost in a benchmark between \physher\ and \phylostan\ \citep{Fourment2019-fh}.
The gradient of the Jacobian transform of the node height reparameterization \citep{Fourment2019-fh} is efficiently calculated using the method proposed by  \citet{Ji2021-hc}.

\begin{table}
\centering
\begin{tabular}{ |c|c| }
 \hline
 Type & Model \\
 \hline
Nucleotide model & JC69, HKY, GTR and SRD06 \\
Amino acid model & WAG and LG \\
%Codon model& GY94 \\
Rate heterogeneity across site & Proportion of invariant sites, Weibull, gamma \\
Phylogeography & Discrete \\
%birth death & constant, BDSKY \\
Coalescent & Constant, skyride, skygrid, skyglide \\
Clock rate & Clock-free, strict, one rate per branch \\
 \hline
\end{tabular}
\caption{Models implemented in \texttt{torchtree-physher}.}
\label{tab:physher-models}
\end{table}

\subsubsection*{torchtree-bito}

\texttt{torchtree-bito} is a \torchtree\ plug-in that offers an interface to the \bito\ library \url{https://github.com/phylovi/bito}.
Within \bito, analytical derivatives with respect to the branch lengths and the parameter of the Weibull site model are calculated through the \BEAGLE\ library \citep{Ayres2019-rw,Ji2020-pp}.
The gradient with respect to the GTR substitution model parameters are calculated numerically using finite differences.
\BEAGLE\ also provides efficient calculation of the tree likelihood and its gradient on GPUs \citep{gangavarapu2024many}.

\subsubsection*{torchtree-scipy and torchtree-tensorflow}
The discretized gamma distribution is widely used to model rate heterogeneity across sites.
The discretization method requires the inverse CDF of the gamma distribution which, at the time of writing (version 2.2.1), is not available in PyTorch.
\texttt{torchtree-scipy} and \texttt{torchtree-tensorflow} are simple plug-ins that implement the gamma-distributed site model using the SciPy and Tensorflow  libraries respectively.
While \texttt{torchtree-tensorflow} calculates gradients using automatic differentiation, \texttt{torchtree-scipy} approximates them using central finite differences.

\subsection*{A continuous piecewise-linear coalescent model}

We have introduced a coalescent model analogous to the piecewise-constant model on a fixed grid (a.k.a. skygrid) \citep{gill2013improving} but instead with a piecewise-linear ancestral population size.
The model, which we call ``skyglide'', is parameterized in terms of ancestral population sizes at times on a fixed grid, and interpolates between them using linear functions.
Specifically, we take

\begin{align*}
p(t_2, \dots, t_{n+1} \mid N(t)) &= \prod_{k=2}^{n} p(t_{k} \mid t_{k+1}, N(t))\\
&= \prod_{k=2}^n {k \choose 2} \frac{1}{N(t_{k})} \exp \left[ -\int_{t_{k+1}}^{t_k} {k \choose 2} \frac{1}{N(t)} dt \right]
\end{align*}

Let $\bm\theta = (\theta_0,  \dots, \theta_M)$ be the vector of effective population sizes at fixed and equidistant time points $0=x_0, \dots, x_M=C$ where $C$ is the user-defined cutoff value of the grid.

We define the piecewise-linear demographic function,

\begin{equation*}
\widehat N(t) =
\begin{cases}
\theta_i + (\theta_{i+1} - \theta_i) \frac{t - x_i}{x_{i+1} - x_i} & \text{if } x_i \leq t \leq x_{i+1} \\
\theta_M & \text{if } t > x_M
\end{cases}.
\end{equation*}

This is in contrast to the piecewise-constant function which assumes $\widehat N_c(t) = \theta_i$ for $x_i \leq t < x_{i+1}$ and $\widehat N_c(t) = \theta_M$ for $t > x_M$.

Each sub-function of $\widehat N(t)$ and $\widehat N_c(t)$ is continuously differentiable since it is either a linear or constant function.
The key difference between the two piecewise functions is that $\widehat N_c(t)$ contains jump discontinuities while $\widehat N(t)$ is continuous across its domain (see section \ref{section:continuity} in Supplementary Material).
Although neither of the piecewise functions is differentiable, our analyses indicate that, unlike the skygrid model, the skyglide model can be effectively utilized with gradient-based algorithms.

\subsection*{Datasets and validation}
We analyzed two datasets to exemplify common and new features implemented in \torchtree, as well as to illustrate the behavior of various objectives for different types of models.

The first dataset comprises 63 RNA sequences of type 4 from the E1 region of the hepatitis C virus (HCV) genome that were isolated in 1993. As in previous studies \citep{pybus2001epidemic}, the substitution rate was fixed to $7.9 \times 10^{-4}$ substitutions per site per year.
As \torchtree\ can only accommodate a single topology, we also enforce the constraint of a fixed tree topology in all BEAST analyses.
The topology used in the \torchtree\ and BEAST analyses was drawn randomly from a sample of trees generated by a preliminary analysis with BEAST without topological constraints.
We used the GTR substitution model and gamma distributed rate heterogeneity with four categories.
For this analysis \texttt{torchtree-physher} plug-in was used to compute gradients for the tree likelihood.
We either used a piecewise-constant (aka skygrid) \citep{gill2013improving} or piecewise-linear (skyglide) population size coalescent prior with a cutoff of 400 years and 75 time segments.
As is customary for piecewise-constant coalescent models \citep{minin2008smooth,gill2013improving} we place a Gaussian Markov random field (GMRF) prior on the vector of log effective population sizes and a gamma prior with rate and scale equal to 0.005 on the precision parameter.
For unconstrained optimization, node heights are reparameterized using the approach outlined by \citet{Fourment2019-fh}.
We used two criteria to estimate the mean-field approximations: ELBO and $\KL(p||q)$ and optimized them for 10 million iterations.
We also performed Hamiltonian Monte Carlo inference with \torchtree\ to approximate the model with the skyglide model for 50 million iterations.
The step size of the leap frog integrator and the diagonal mass matrix were tuned automatically.
Every model was approximated using 50 million MCMC iterations in BEAST.

The second dataset is made of 583 SARS-CoV-2 RNA sequences from \citep{pekar2021timing}.
Using a fixed topology, we replicated the analyses of \citet{magee2023random} to assess whether there is a rate increase of C$\rightarrow$T substitutions over the reverse T$\rightarrow$C substitutions.
The rooted topology is obtained through a two-step process: first, by estimating the maximum likelihood (unrooted) tree using iqtree \citep{minh2020iqtree}, and then by determining the root location using lsd \citep{to2016lsd}.
We implemented the HKY substitution model with random effects to allow for nonreversibility (see \citet{magee2023random} for more information on the model).
Due to discontinuities in the skygrid model we used a skyglide model with 5 parameters and a cutoff of 0.3 years.
Although \citet{magee2023random} used a regularized Bayesian bridge prior \citep{nishimura2023bridge}, we opted to use the original Bayesian bridge formulation \citep{polson2013bridge}.
The Bayesian bridge prior on random effect $\epsilon$ has density
$$p(\epsilon | \tau, \alpha) \propto \exp\left(-\left|\frac{\epsilon}{\tau}\right|^\alpha \right).$$

In this study, the exponent is fixed to $\alpha=0.25$ and we place a gamma prior on $\tau^{-\alpha}$ with shape $\delta=1$ and scale $\theta=2$.

We can test the support for nonreversibilities, for example the difference between the C$\rightarrow$T and T$\rightarrow$C rates, with Bayes factors.
The fact that a model with the C$\rightarrow$T and T$\rightarrow$C rates equal (reversible with respect to C$\leftrightarrow$T) is nested within the random-effects model allows us to use the Savage-Dickey ratio (see \citep{wagenmakers2010bayesian} for an example) to compute the Bayes factor from the posterior distribution of the random-effects model.

As in the original study, we used the GTR substitution model to visually inspect the effect of using random effects in the substitution rate matrix.
The GTR and HKY-RE models were approximated using 50 million MCMC iterations in BEAST.
For the variational inference analyses, we optimized the ELBO for 1 million iterations.

In our analyses, we employed the mean-field approximation, modeling each factor as a univariate Gaussian distribution.
To ensure that the phylogenetic parameters adhered to their respective constraints, we applied differentiable invertible transformations, aligning them with the support of the Gaussian distributions.
For example, we used the exponential function to transform positive parameters, and the logistic function to map parameters in the interval $[0,1]$ onto the real line.
These transformations, along with the mean-field approximation, are the foundations of the automatic differentiation variational inference framework \citep{Carpenter2017-ui,Kucukelbir2017-cr}.

We investigated the speed of convergence of MCMC and variational inference using the coefficient of variation (CV) of the approximated distributions.
We calculated the CV of each distribution at multiple time points and compared them to the CV of the final approximations.
For an analysis of $M$ iterations taking $t_M$ units of time, a time point indexed by its iteration number $i$ is defined as $t_i=i \times t_M / M$.
For MCMC, the coefficient of variation $CV(t_i)$ is calculated using samples from iteration $1$ up to iteration $i$.
With variational inference, we calculate the smoothed mean $\bar\mu(t_i)$ and variance $\bar\sigma^2(t_i)$ using the variational distributions from time $1$ up to time $t_i$.
We then compute the coefficient of variation $CV(t_i) = \sqrt{\bar\sigma^2(t_i)} / \bar\mu(t_i)$ where $\bar\mu(t_i) = \frac{\sum_{j=1}^i \mu(t_j)}{i}$ and $\bar\sigma^2(t_i) = \frac{\sum_{j=1}^i \sigma^2(t_j)}{i}$.
In the corresponding plots, we show the relationship between $t_i$ for $\{i\times 1000 | i=1,2,3,\dots,M/1000\}$ and the squared difference $(CV(t_i) - CV(t_M))^2$, showing how quickly the algorithms convergence over time.

Given that the \textit{a priori} choice of the number of iterations resulted in different run times for BEAST and \torchtree, we use the same total runtime (i.e. $t_M$) for both.
This approach is equivalent to having a consistent time budget constraint across the two methods.

In the interest of reproducibility, a Nextflow pipeline running every analysis is available from \url{https://github.com/4ment/torchtree-experiments}.

\section*{Results}

\subsection*{Effective population size estimation of HCV dataset}
\begin{figure*}[h]
\centering
\includegraphics[width=0.95\textwidth]{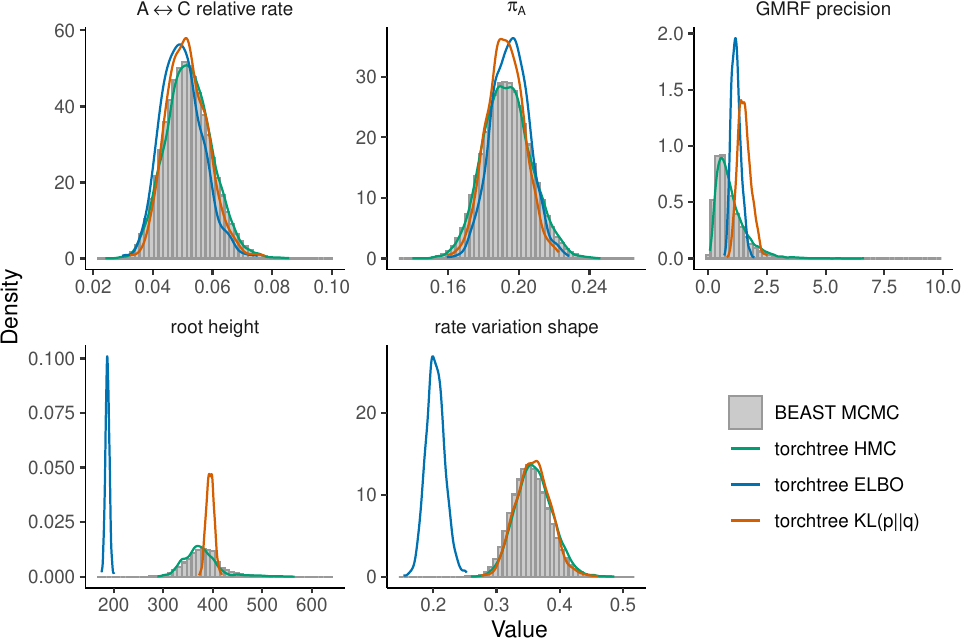}
\caption{\
Posterior approximation of phylogenetic model parameters using \torchtree\ and BEAST on the HCV dataset with the skygrid (piecewise-constant) model.
\torchtree\ approximates the distributions using either mean-field variational inference (ELBO and $\KL(p||q)$) or MCMC.
BEAST uses MCMC.
The plot displays density distributions for several parameters: the substitution rate bias between nucleotide A and C ($\text{A} \leftrightarrow \text{C}$), the frequency of nucleotide A ($\pi_{\text{A}}$), the GMRF precision parameter, the age of the root node (root height) and the shape parameter of the discrete gamma site model.
The gradient-based ELBO inference clearly struggles in this case of a discontinuous model.
}%
\label{fig:skygrid-params}
\end{figure*}

\begin{figure*}[h]
\centering
\includegraphics[width=0.95\textwidth]{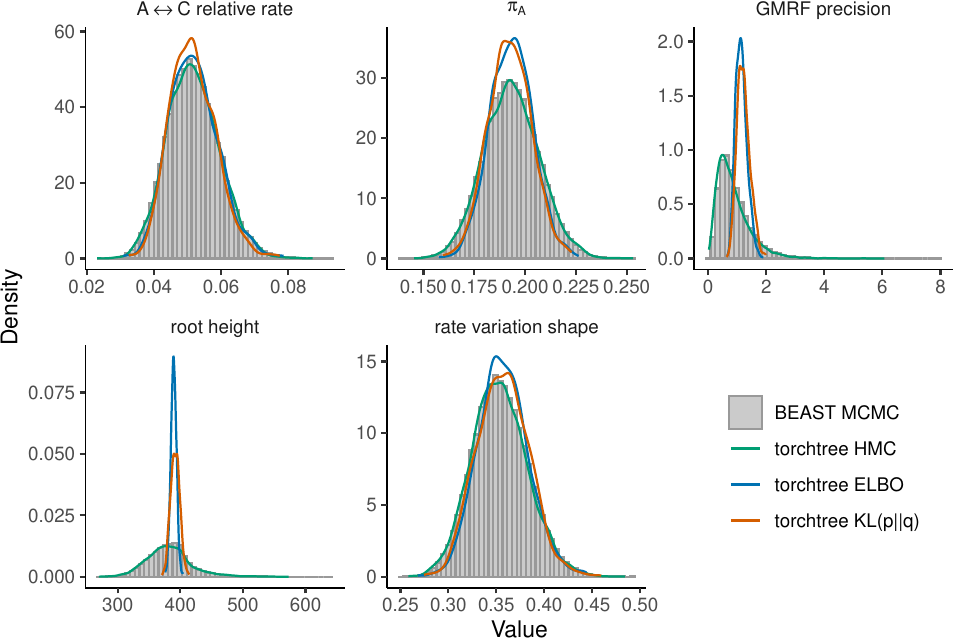}
\caption{\
Posterior approximation of phylogenetic model parameters using \torchtree\ and BEAST on the HCV dataset with the skyglide (piecewise-linear) model.
\torchtree\ approximates the distributions using either mean-field variational inference (ELBO and $\KL(p||q)$) or Hamiltonian Monte Carlo (HMC).
BEAST uses MCMC.
The plot displays density distributions for several parameters: the substitution rate bias between nucleotide A and C ($\text{A} \leftrightarrow \text{C}$), the frequency of nucleotide A ($\pi_{\text{A}}$), the GMRF precision parameter, the age of the root node (root height) and the shape parameter of the discrete gamma site model.
}%
\label{fig:skyglide-params}
\end{figure*}

First, we analyzed a dataset consisting of HCV RNA sequences under two piecewise coalescent priors: piecewise-constant (skygrid) and piecewise-linear (skyglide) models.
When utilizing the ELBO criterion and gradient-based optimization, it becomes evident that the discontinuous piecewise-constant model struggles to retrieve the root height and shape parameter of the gamma site model (Figure~\ref{fig:skygrid-params}), unlike the piecewise-linear model which offers more accurate approximations (Figure~\ref{fig:skyglide-params}).
In contrast, optimizing the $\KL(p||q)$ objective yields more accurate approximations of these parameters under both piecewise models.
Since the $\KL(p||q)$ objective does not require the calculation of the gradient of the coalescent, this suggests that gradient computations may be at fault.
The existence of jump discontinuities in the piecewise-constant model makes it infeasible to employ gradient-based optimization methods.
By construction, the piecewise-linear model used in this study is continuous since the endpoint of one segment is the initial point of the next segment.
The $\KL(p||q)$-based analysis was significantly faster than its ELBO counterpart, as the former method required 300 minutes, while the latter took 930mins.

\begin{figure*}[h]
\centering
\includegraphics[width=0.95\textwidth]{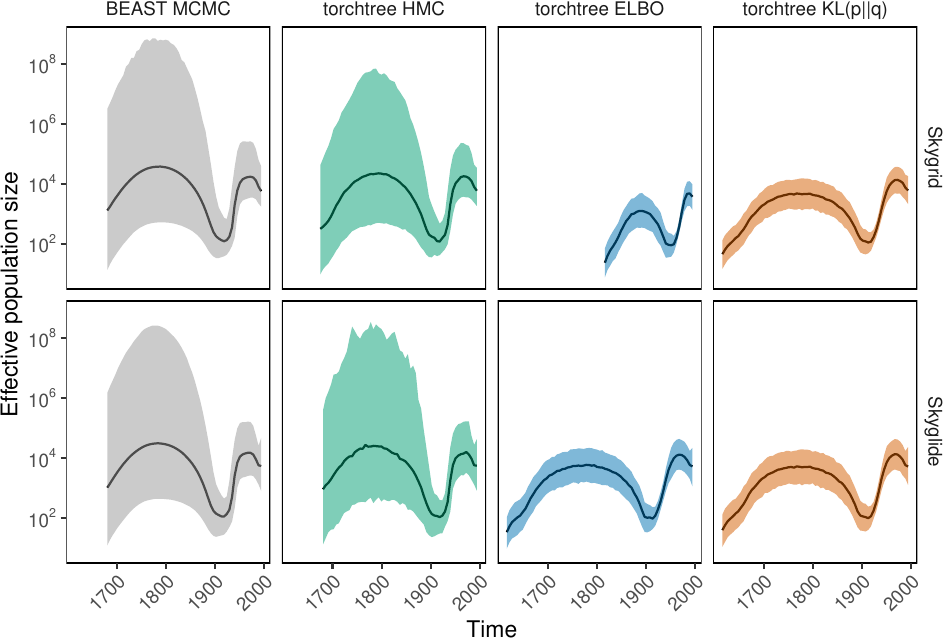}
\caption{\
Posterior approximation of skyglide (piecewise-linear) and skygrid (piecewise-constant) distributions using \torchtree\ and BEAST on the HCV dataset.
\torchtree\ approximates the distributions using mean-field variational inference (ELBO and $\KL(p||q)$) and HMC.
}%
\label{fig:sky-plots}
\end{figure*}

\subsection*{C to T bias in SARS-CoV-2 evolution}
\begin{figure*}[h]
\centering
\includegraphics[width=0.95\textwidth]{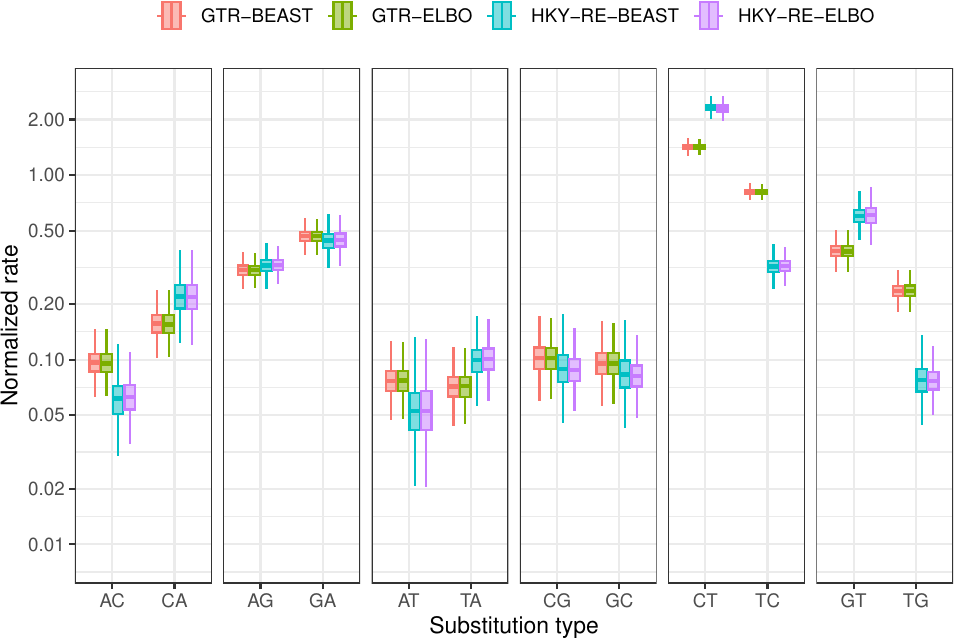}
\caption{\
Posterior distributions of the 12 non-diagonal elements of the inferred rate matrices for the dataset of \citet{magee2023random} using BEAST (MCMC) and \torchtree\ (variational inference with ELBO).
The solid line is the posterior median, the shaded region the 50\% CI.
The whiskers extend to the posterior samples farthest from the median but within $1.5\times$ the interquartile range.
}%
\label{fig:sc2}
\end{figure*}

We find that the normalized substitution rates estimated using VB with \torchtree\ closely match those obtained through BEAST (Figure~\ref{fig:sc2}).
Even though \citet{magee2023random} sampled the full topology space, we also find evidence for a greatly elevated rate of C$\rightarrow$T substitutions, as well as an elevated G $\rightarrow$T rate.
The Bayes factor provides “very strong” [Kass and Raftery, 1995] support for the nonreversibility of C$\rightarrow$T and G$\rightarrow$T rates (over the reversible model) (Table~\ref{table:sc2}).

The ELBO-based and $\KL(p||q)$-based approximations for the rate parameters of the substitution matrix were similar for both the GTR (Supplementary Figure~\ref{fig:sc2-gtr-params}) and HKY-RE (Supplementary Figures~\ref{fig:sc2-hkyre-params} and \ref{fig:sc2-hkyre-coefs}) models.
However, the VB approximations using the $\KL(p||q)$ criterion showed slightly poorer performance when approximating some of the population size parameters (Supplementary Figure~\ref{fig:sc2-gtr-coalescent}).

Although we expected VB to be significantly faster than MCMC, we found that this was not the case.
To investigate this question, we examined how the coefficient of variation evolves over time relative to the final approximation (Supplementary Figures~\ref{fig:sc2-hkyre-cv-params} and \ref{fig:sc2-hkyre-cv-coalescent}).
We found that this CV converged at similar rates for the MCMC-based and VB-based analyses, although the latter is significantly more ``jagged'' because it represents a point estimate of the variance rather than an averaged variance.
These results also suggest that ELBO-based inference tend to converge more quickly than the $\KL(p||q)$-based analysis, even though the former is more computationally intensive due to gradient calculations.

\begin{table}[h!]
\centering
\begin{tabular}{| c | c | c| }
\hline
Substitution & \multicolumn{2}{|c|}{Log Bayes factor} \\
 & torchtree & BEAST \\ \hline
A$\leftrightarrow$C & 0.9 & 0.6 \\ \hline
A$\leftrightarrow$G & -0.6 & -0.57 \\ \hline
A$\leftrightarrow$T & 0.52 & 0.4 \\ \hline
C$\leftrightarrow$G & -0.46 & -0.43 \\ \hline
C$\leftrightarrow$T & 25.9 & 22.08 \\ \hline
G$\leftrightarrow$T & 14.2 & 8.16 \\ \hline
\end{tabular}
\caption{Bayes factors favoring nonreversibility ($\BF_{10}$) calculated using the Savage-Dickey ratio.}
\label{table:sc2}
\end{table}

\section*{Discussion and Conclusion}

The presented study introduces \torchtree, a novel program for phylogenetic inference utilizing variational inference and other gradient-based algorithms.
\torchtree\ exhibits notable advantages, including enhanced speed compared to other automatic differentiation-based tools like \phylostan, and provides the capability to formulate and implement complex models and algorithms.

In our results we highlighted that the piecewise-constant coalescent model, a non-differentiable and discontinuous function, can lead to spurious results with variational inference.
Although the HMC analyses did not show similar problems for the HCV dataset, there is no guarantee that analyses with other datasets will not encounter issues.
We advocate the use of the piecewise-linear model since it has similar complexity and expressiveness to the piecewise-constant model, yet its continuous structure allows for gradient-based variational Bayesian methods.

Our results showed that gradient-free variational inference with $\KL(p||q)$ was fast and accurate, especially for parameters of the substitution model for which gradient are notoriously expensive \citep{fourment2023automatic,magee2023random}.
To accelerate phylogenetic variational inference, a promising direction for research would be to optimize the $\KL(p||q)$ objective for these parameters and maximize the ELBO for the other parameters.

It is important to note that, at present, \torchtree\ necessitates a pre-existing phylogenetic tree topology as input and the topology remains fixed throughout the entire inference process.
While this limitation simplifies certain aspects of the analysis, ongoing research endeavors are actively exploring ways to incorporate topology space exploration \citep{Zhang2022-tr,Koptagel2022-bk}.
The inclusion of such functionality would represent a substantial enhancement, allowing for a more comprehensive exploration of phylogenetic relationships.

ELBO-based inference converged more quickly to the final approximation, suggesting that incorporating gradient information in the optimization problem was beneficial, despite its computational complexity.
Although variational inference tends to be seen as a faster alternative to MCMC, our results did not support this statement.
VB appeared to approximate some parameters, such as the root height, more slowly than MCMC.

Another direction for future work is to extend beyond the mean-field variational inference used here, which is a fully factorized variational family that ignores correlation among latent variables.
A substantial body of research model dependencies between latent variables through normalizing flows, a framework relying on a series of transformations typically learned with neural networks \citep{rezende2015normalizing}.
Although \citet{zhang2020normalizing} reported some promising results, \citet{Ki2022-nt} suggested that normalizing flows did not yield improvements in the approximation.
This discrepancy may stem from the substantial differences in the phylogenetic models explored by these researchers.
Our preliminary analyses with normalizing flow have also showed no improvement in the final approximation (data not shown).
A deeper investigation into normalizing flows is warranted, as well as exploration of other methods that build structured variational families \citep{yin2018semiimplicit,ambrogioni2021structured}.

\section*{Acknowledgments}
MM and MF were supported by the Australian Government through the Australian Research Council (project number LP180100593).
XJ acknowledges support through Louisiana Board of Regents Research Competitiveness Subprogram and NSF grant DEB1754142.
This project was partially supported by US National Institutes of Health grants R01 AI162611 and R01 AI153044.
Computational facilities were provided by the UTS eResearch High Performance Computer Cluster.
%
%This work was supported through US National Institutes of Health grant AI162611.
Scientific Computing Infrastructure at Fred Hutch was funded by ORIP grant S10OD028685.
Dr.\ Matsen is an Investigator of the Howard Hughes Medical Institute.

\clearpage

\bibliographystyle{plainnat}
\bibliography{main}

\beginsupplement
\clearpage

%\subfile{SI}
%\documentclass{article}
%\documentclass[main.tex]{subfiles}

% \usepackage{graphicx}
% \usepackage{amsmath}

\ifdefined\KL
\renewcommand{\figurename}{Supplementary Figure}
\else
\newcommand{\KL}{\operatorname{KL}}
\newcommand{\torchtree}{\texttt{torchtree}}
\fi

%\renewcommand{\thefigure}{S\arabic{figure}}
%\renewcommand{\thepostfigure}{S\arabic{postfigure}}

% \begin{document}
\begin{center}
\textbf{Supplementary Material}
\end{center}

\section{Methods}

\subsection{Continuity and differentiability of piecewise-linear coalescent model} \label{section:continuity}
A piecewise-linear function consists of multiple linear segments joined at specific points, known as "breakpoints." Although each segment is linear, the overall function is continuous if it does not have any abrupt jumps at the breakpoints.
In a continuous function, the limit from the left (as you approach a point from the left side) and the limit from the right (as you approach a point from the right side) must be equal to the value of the function at that point.

For $M \geq 3$ segments, two adjacent segments that do not include the last segment are defined as

\begin{equation*}
\widehat N(t) =
\begin{cases}
    \theta_i + (\theta_{i+1} - \theta_i) \frac{t - x_i}{x_{i+1} - x_i} & \text{if } x_i \leq t \leq x_{i+1} \\
    \theta_{i+1} + (\theta_{i+2} - \theta_{i+1}) \frac{t - x_{i+1}}{x_{i+2} - x_{i+1}} & \text{if } x_{i+1} \leq t \leq x_{i+2} \\
\end{cases}.
\end{equation*}

These segments are continuous since $\lim_{t\to x_{i+1}^-} \widehat N(t) = \lim_{t\to x_{i+1}^+} \widehat N(t) = \widehat N(x_{i+1}) = \theta_{i+1}$.

For the last two segments we have

\begin{equation*}
\widehat N(t) =
\begin{cases}
    \theta_{M} + (\theta_{M} - \theta_{M-1}) \frac{t - x_{M-1}}{x_{M} - x_{M-1}} & \text{if } x_{M-1} \leq t \leq x_{M} \\
    \theta_{M} & t \geq \text{if } x_{M}  \\
\end{cases}.
\end{equation*}

The last two segments are also continuous since $\lim_{t\to x_{M}^-} \widehat N(t) = \lim_{t\to x_{M}^+} \widehat N(t) = \widehat N(x_{M}) = \theta_{M}$.

The piecewise-linear model is continuous since $\lim_{t\to x_{i+1}^-} \widehat N(t) = \lim_{t\to x_{i+1}^+} \widehat N(t) = \widehat N(x_{i+1})$ at every boundary.

A function is differentiable at a point if it has a well-defined, finite derivative at that point. For a piecewise-linear function, differentiability requires that the function be both continuous and have matching derivatives from the left and right at each breakpoint.

\begin{equation*}
\widehat N'(t) =
\begin{cases}
    \frac{\theta_{i+1} - \theta_i}{x_{i+1} - x_i} & \text{if } x_i \leq t \leq x_{i+1} \\
    \frac{\theta_{i+2} - \theta_{i+1}}{x_{i+2} - x_{i+1}} & \text{if } x_{i+1} \leq t \leq x_{i+2} \\
\end{cases}.
\end{equation*}

The function is not differentiable since $\lim_{t\to x_{i+1}^-} \widehat N'(t) \neq \lim_{t\to x_{i+1}^+} \widehat N'(t)$.

\subsection{Bayes factor calculation for meanfield variational inference}

Following the notation of \citet{magee2023random}, the Bayes factor in favor of $\Delta_{ij} = 0$ (Model 0, against Model 1 where $\Delta_{ij}$ is a free parameter) is the ratio of the posterior density to the prior density at $\Delta_{ij} = 0,$

\[
\BF_{01} = \frac{p(\Delta_{ij}=0 | y)}{p(\Delta_{ij}=0)} \approx \frac{q^{*}(\Delta_{ij}=0)}{p(\Delta_{ij}=0)},
\]

where $q^{*}$ is a normal distribution with mean $m_{ji} - m_{ij}$ and variance $\sigma_{ij}^2 + \sigma_{ji}^2$.
$m_{ij}$ and $\sigma_{ij}$ are the mean and standard deviation of the normal variational approximation of $\epsilon_{ij}$.
The Bayes factor favoring nonreversibility ($\Delta_{ij} \neq 0$) is $\BF_{10} = 1/\BF_{01}$.
\citep{magee2023random} provide a closed-form expression for the probability density of $\Delta_{ij}=0$ evaluated at 0

$$p(\Delta_{ij}=0) = \frac{\alpha \Gamma(\delta + 1/\alpha)}{2^{1+1/\alpha} \Gamma(1/\alpha)\Gamma(\delta)\beta^{1/\alpha}}.$$

\section{Results}

\begin{figure*}[h]
\centering
\includegraphics[width=0.95\textwidth]{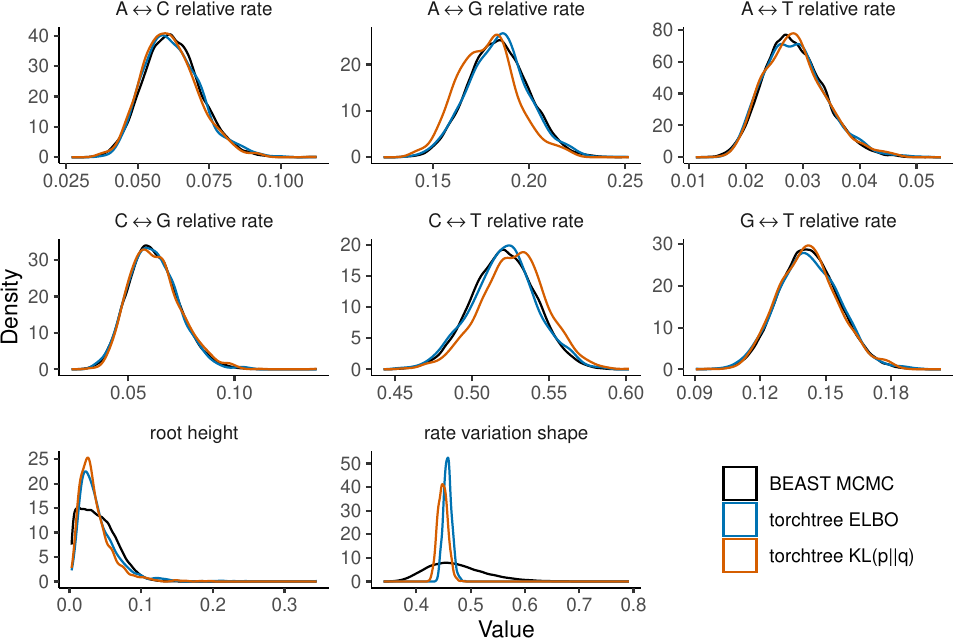}
\caption{\
Posterior approximation of phylogenetic model parameters using \torchtree\ and BEAST on the SC2 dataset with the GTR substitution model.
\torchtree\ approximates the distributions using variational inference  with either the ELBO or $\KL(p||q)$ objective functions, whereas BEAST employs MCMC.
The plot displays density distributions for several parameters: the substitution rate bias parameters ($\text{A} \leftrightarrow \text{C}, \text{A} \leftrightarrow \text{G}, \dots, \text{G} \leftrightarrow \text{T}$), the age of the root node (root height) and the shape parameter of the discrete gamma site model.
}%
\label{fig:sc2-gtr-params}
\end{figure*}

\begin{figure*}[h]
\centering
\includegraphics[width=0.95\textwidth]{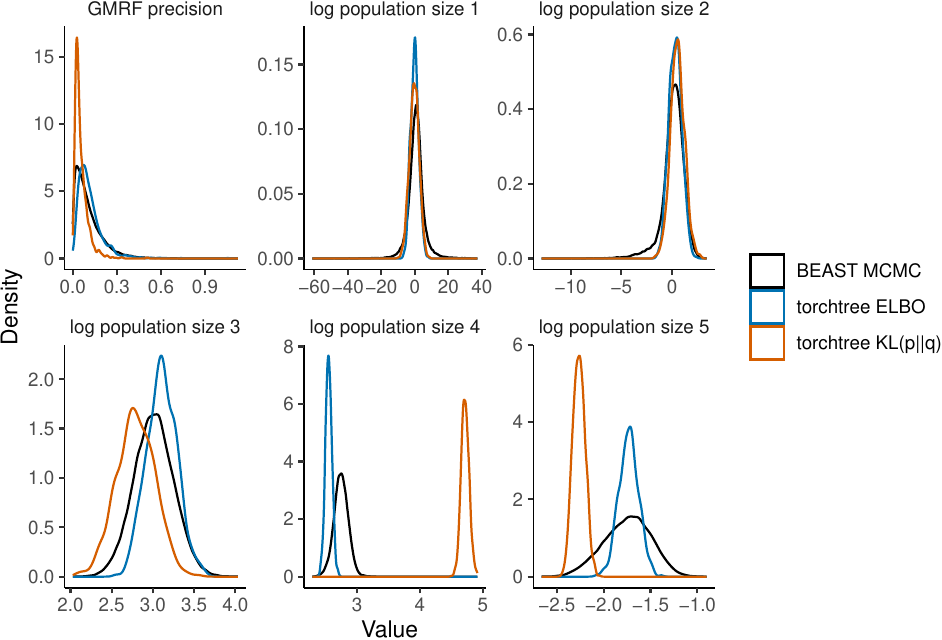}
\caption{\
Posterior approximation of the parameters of the coalescent and Gaussian Markov random field (GMRF) models using \torchtree\ and BEAST on the SC2 dataset with the GTR substitution model.
\torchtree\ approximates the distributions using variational inference  with either the ELBO or $\KL(p||q)$ objective functions, whereas BEAST employs MCMC.
The plot displays density distributions for the five population size parameters and the precision parameter of the GMRF.
}%
\label{fig:sc2-gtr-coalescent}
\end{figure*}

\begin{figure*}[h]
\centering
\includegraphics[width=0.95\textwidth]{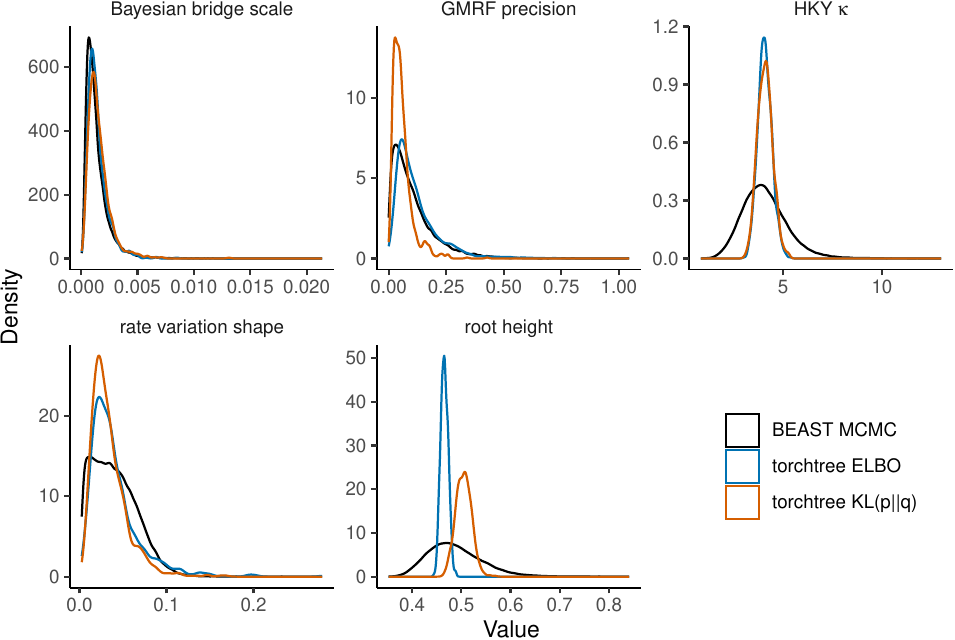}
\caption{\
Posterior approximation of phylogenetic model parameters using \torchtree\ and BEAST on the SC2 dataset with the HKY-RE substitution model.
\torchtree\ approximates the distributions using variational inference  with either the ELBO or $\KL(p||q)$ objective functions, whereas BEAST employs MCMC.
The plot displays density distributions for several parameters: the scale parameter of the Bayesian bridge prior, the precision parameter of the GMRF, the ratio of transition and transversion rate parameter ($\text{HKY} \kappa$) of the HKY model, the shape parameter of the discrete gamma site model,  and the age of the root node.
}%
\label{fig:sc2-hkyre-params}
\end{figure*}

\begin{figure*}[h]
\centering
\includegraphics[width=0.95\textwidth]{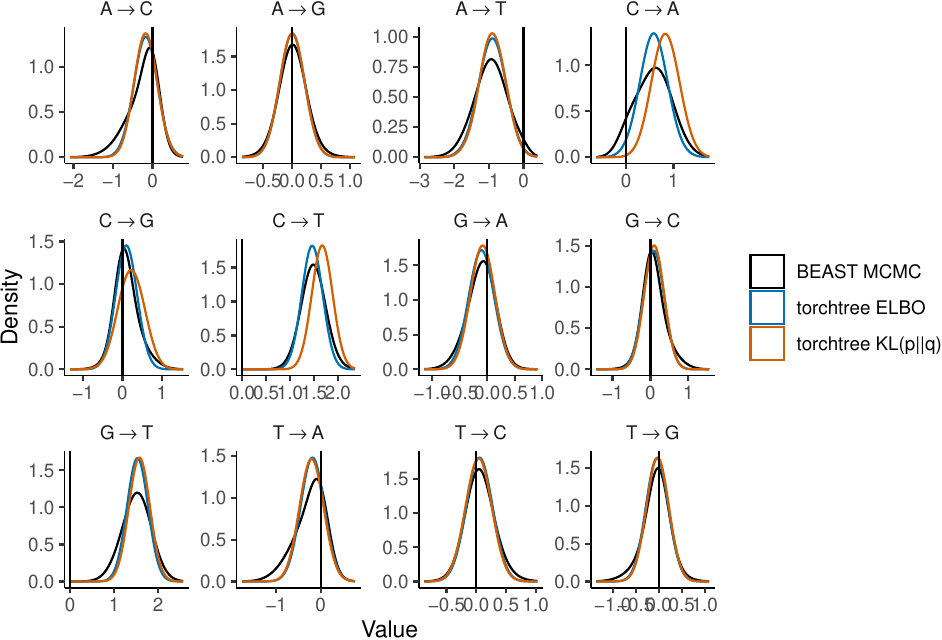}
\caption{\
Posterior approximation of the random effect parameters using \torchtree\ and BEAST on the SC2 dataset with the HKY-RE substitution model.
\torchtree\ approximates the distributions using mean-field variational inference (ELBO and $\KL(p||q)$) and BEAST uses MCMC.
}%
\label{fig:sc2-hkyre-coefs}
\end{figure*}

\begin{figure*}[h]
\centering
\includegraphics[width=0.95\textwidth]{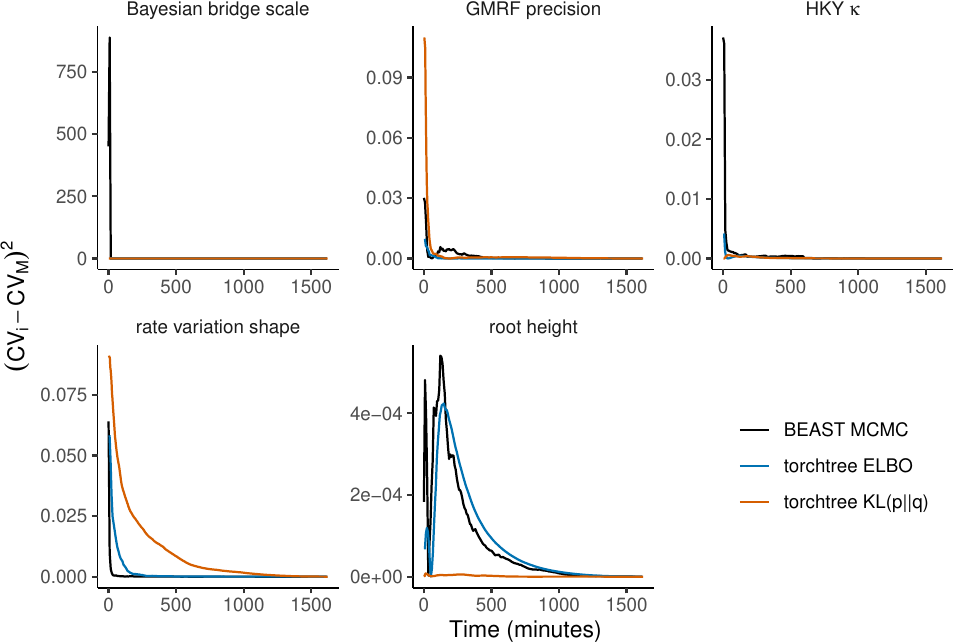}
\caption{\
Convergence between intermediate estimates and the final approximation of phylogenetic model parameters using \torchtree\ and BEAST on the SC2 dataset with the HKY-RE substitution model.
\torchtree\ approximates the distributions using variational inference  with either the ELBO or $\KL(p||q)$ objective functions, whereas BEAST employs MCMC.
The y-axis represents the squared difference between the coefficient variation (CV) at time $i$ and the CV of the final approximation at time $M$ while the x-axis represent time.
}%
\label{fig:sc2-hkyre-cv-params}
\end{figure*}

\begin{figure*}[h]
\centering
\includegraphics[width=0.95\textwidth]{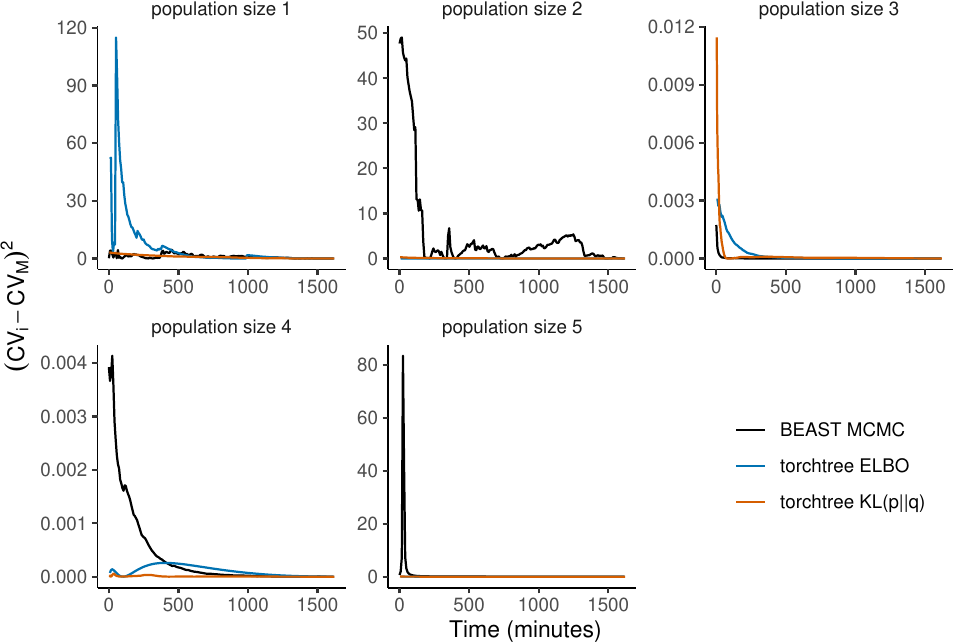}
\caption{\
Convergence between intermediate estimates and the final approximation of the five population size parameters using \torchtree\ and BEAST on the SC2 dataset with the HKY-RE substitution model.
\torchtree\ approximates the distributions using variational inference  with either the ELBO or $\KL(p||q)$ objective functions, whereas BEAST employs MCMC.
The y-axis represents the squared difference between the coefficient variation (CV) at time $i$ and the CV of the final approximation at time $M$ while the x-axis represent time.
}%
\label{fig:sc2-hkyre-cv-coalescent}
\end{figure*}

% \end{document}

\end{document}